\documentclass[pra,twocolumn,superscriptaddress,aps,showpacs,amsmath,amssymb,nofootinbib]{revtex4}
\usepackage{graphicx}
\usepackage{dcolumn}
\usepackage{bm}
\newcommand{\bce}{\begin{center}}
\newcommand{\ece}{\end{center}}
\newcommand{\beq}{\begin{equation}}
\newcommand{\eeq}{\end{equation}}
\newcommand{\beqa}{\begin{eqnarray}}
\newcommand{\eeqa}{\end{eqnarray}}
\begin{document}

\title{Spinor BECs in a double-well: population transfer and Josephson oscillations }
\date{\today}

\author{B. Juli\'a-D\'\i az} 
\affiliation{Departament d'Estructura i Constituents de la Mat\`{e}ria,\\
Universitat de Barcelona, 08028 Barcelona, Spain}

\author{M. Mel\'{e}-Messeguer}
\affiliation{Departament d'Estructura i Constituents de la Mat\`{e}ria,\\
Universitat de Barcelona, 08028 Barcelona, Spain}

\author{M. Guilleumas}
\affiliation{Departament d'Estructura i Constituents de la Mat\`{e}ria,\\
Universitat de Barcelona, 08028 Barcelona, Spain}

\author{A. Polls}
\affiliation{Departament d'Estructura i Constituents de la Mat\`{e}ria,\\
Universitat de Barcelona, 08028 Barcelona, Spain}

\begin{abstract}
The dynamics of an $F=1$ spinor condensate in a two-well potential 
is studied within the framework of the Gross-Pitaevskii equation. 
We derive two-mode equations relating the population imbalances, the 
phase differences among the condensates at each side of the barrier and the 
time evolution of the different Zeeman populations for the case of 
small population imbalances. The case of zero total magnetization 
is scrutinized in this limit demonstrating the ability of a two mode 
analysis to describe to a large extent the dynamics observed in the 
Gross-Pitaevskii equations. It is also demonstrated that the time 
evolution of the different total populations fully decouples from 
the Josephson tunneling phenomena. All the relevant time scales are 
clearly identified with microscopic properties of the atom-atom 
interactions.
\end{abstract}


\maketitle

The recent experimental work on optically trapped spinor 
condensates has broadened the frontiers of confined
Bose Einstein condensates (BECs)~\cite{chang1,chang2}. There, 
a transfer of population between the different Zeeman 
components of a spinor BEC was observed providing a 
clear signal of the spin-dependent interatomic interactions. 
This experiment quickly connected the field of cold-atoms 
to a large variety of problems in quantum magnetism, mostly 
related to magnetic ordering and spin dynamics~\cite{lewenstein}. 
From a different point of view, it can be regarded as the first 
case of ternary mixture of BECs with population exchange 
among the three components, a nice example of coupled multicomponent 
quantum gases.
 
At the same time, the fast tunneling of atoms through potential 
barriers driven by imbalanced populations at each side of the 
barrier was shown experimentally, short after it was observed 
in optical lattices~\cite{kasevich}, providing the first 
confirmation of Josephson tunneling of atoms in BECs~\cite{Albiez05}.
From the BEC point of view, Josephson tunneling through the 
potential barrier produces a weak coupling between the BECs 
at each side of the trap, presenting a coupled-oscillator behavior 
in the appropriate variables~\cite{Gati2007}. 

Both effects have been considered by several theoretical 
groups in quite different contexts, e.g.
~\cite{tunnellingatoms,Smerzi97,Gati2007,ananikian,spinorBEC,spin,book}. 
In this letter we address both aspects together: the Josephson 
oscillations and the transfer of populations. Providing a direct 
connection between the different time scales and the microscopic 
properties of the interatomic interactions.

To fix the conditions we consider the simplest scenario, which already 
contains relevant physics. We restrict our analysis to the 
case of small population imbalance of all the Zeeman sublevels and 
small initial phase difference between the same component at 
both sides of the trap. Our main tool are the mean-field Gross-Pitaevskii (GP)  
equations for spinor BECs~\cite{spinorBEC}. On top of numerical 
solutions to the GP equations we illustrate the physics emerging 
in this fairly complex situation by deriving a two-mode description of the 
problem. Within this two mode description it is easy to show that 
for small population imbalances and phase differences the population 
transfer dynamics fully decouples from the tunneling phenomena.

In the mean field approximation the dynamics of the vector 
order parameter $\Psi=(\Psi_{-1},\Psi_0,\Psi_1)$ representing 
the $F=1$ spinor condensate is given by~\cite{spinorBEC}, 
\beqa
\imath \hbar \partial_t \Psi_{\pm 1} &=&
\left[ {\cal H}_s + c_2 (n_{\pm1}+n_0-n_{\mp 1})\right ] \Psi_{\pm 1} 
+ c_2 \Psi_0^2 \Psi_{\mp 1}^*
\nonumber \\
\imath \hbar \partial_t \Psi_{0} &=&
\left[ {\cal H}_s+ c_2 (n_{1}+ n_{-1})\right ] \Psi_{0} + c_2 2 \Psi_1
\Psi_0^* \Psi_{-1}  \,,
\label{eq:gp}
\eeqa
where, ${\cal H}_s=-{\hbar^2 \over 2 M} \nabla^2 + V + c_0 n $, 
$n_m(\vec{r},t)=|\Psi_m(\vec{r},t)|^2$, 
$n(\vec{r},t)=\sum_m n_m (\vec{r},t)$, and $m=0,\pm 1$.  
The population of each hyperfine sublevel is 
$N_{m}(t) = \int d\vec{r} \;n_m(\vec{r},t) \,$. Due to the 
last term in the r.h.s of Eqs.~(\ref{eq:gp}), the population 
of each Zeeman sublevel is not conserved.  
The couplings are $c_0=4\pi\hbar^2(a_0+2a_2)/(3M)$ and 
$c_2=4\pi\hbar^2(a_2-a_0)/(3M)$, where $a_0$ and $a_2$ are the 
scattering lengths describing binary elastic collisions in the channels 
of total spin 0 and 2, respectively. Their values for $^{87}$Rb are 
$a_0=101.8 a_B$ and $a_2=100.4 a_B$\cite{vanKempen}, which 
yield $c_2<0$, thus producing a ferromagnetic-like 
behavior. The total number of atoms in the system and total magnetization 
are conserved quantities, $N =  \int d\vec{r} \; n(\vec{r},t)$
and $M=\int d\vec{r} [n_{+1}(\vec{r},t)-n_{-1}(\vec{r},t)]\,.$

We consider a setup similar to that described in 
Ref.~\cite{Albiez05} but with two important differences: the 
total number of atoms and the barrier height. In our case the 
number of atoms is larger, $N=15000$, in order to enhance 
population transfer effects. We use the same kind of double-well potential 
but with a higher barrier and a tighter confinement in the 
$x$ direction to ensure a clear Josephson tunneling 
situation. The potential then reads, 
\begin{eqnarray}
V(\vec{r})&=&{M\over 2}( \omega_x^2 x^2  +\omega_y^2 y^2 + \omega_z^2 z^2) +V_0 \cos^2(\pi x/q_0) 
\nonumber
\end{eqnarray}
with $\omega_x=2 \pi \times 100$ Hz, $\omega_y=2 \pi \times 66$ Hz,
$\omega_z=2 \pi \times 90$ Hz, $q_0=5.2 \mu$m, $V_0=3500 \,h$ Hz and 
$M$ is the mass of the atoms. As in the experiment~\cite{Albiez05} 
we assume that the dynamics takes place essentially on the $x$ axis. 
Then, defining $\omega_\perp=\sqrt{\omega_z \omega_y}$ the coupling 
constants can be rescaled by a factor $1/(2\pi a_{\perp}^2)$, with 
$a_{\perp}$ the transverse oscillator length~\cite{Moreno2007}, and the dynamical 
equations transform to one-dimensional ones for a symmetric double-well. 

\begin{figure}[t]
\includegraphics[width=0.95\columnwidth, angle=0, clip=true]{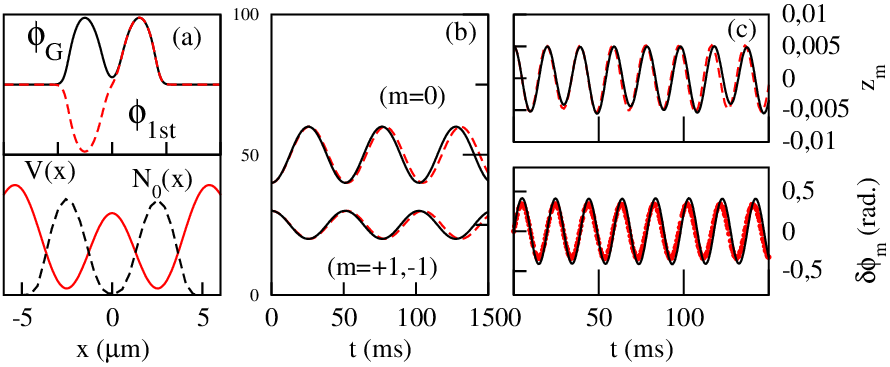}
\caption[]{
(a) (above) $\Phi_G$ and $\Phi_{1st}$. (below) Potential 
in the $x$ direction together with one of the initial 
population profiles used in the simulations (arbitrary units).
The solid (black) curves show the evolution of the total population 
(b) and of the population imbalances and phase differences (c) 
corresponding to simulation I of Tab.~\ref{tab:1}. The dashed (red)
lines depict the two-mode calculation. }
\label{fig1}
\end{figure}

The numerical simulations of Eqs.~(\ref{eq:gp}) are performed in the 
following way. First using an imaginary time evolution method we compute 
the ground, $\Phi_{GS}\equiv \Phi_+$, and first excited state, 
$\Phi_{1st}\equiv \Phi_-$ of a scalar BEC, $c_2=0$, under the same conditions. Then, 
given initial population imbalances for all the components, we build $t=0$ 
wave functions by the appropriate linear combinations of $\Phi_+$ and 
$\Phi_-$. We study the time evolution of the system by means 
of the split operator method. In Fig.~\ref{fig1} we depict 
$\Phi_{GS}$, $\Phi_{1st}$ and the potential in the $x$ direction 
together with one of the initial density profiles used in the simulations.

To characterize the Josephson dynamics we define for each component the 
population imbalance, $z_m=(N_{m,L}-N_{m,R})/N_m$, and the phase difference, 
$\delta\phi_m=\phi_{m,R}-\phi_{m,L}$. Where, $N_{m,L}(t)=
\int_{-\infty}^0 {\rm d} x\,\int_{-\infty}^{\infty}\int_{-\infty}^{\infty}
{\rm d} y\,{\rm d} z\,n_m(\vec{r},t)$, 
$N_{m,R}(t)=N_m(t)-N_{m,L}(t)$ and $\phi_{m,R(L)}$ are
the space average of the phase of $\Psi_m(\vec{r},t)$ at each side of the 
barrier. Lets us emphasize that the phase of $\Psi_m(\vec{r},t)$ is almost 
spatially constant at each side of the trap during the GP simulations.

To better understand the dynamical content of the GP results,
Eqs.~(\ref{eq:gp}), we derive a two mode approximation for the 
GP description of spinor BEC based on the 
following ansatz, 
$
\Psi_{m}(\vec{r},t) = \Psi_{m,L}(t) \Phi_{m,L}(\vec{r}) +
\Psi_{m,R}(t)
\Phi_{m,R}(\vec{r})\,,
$
where the modes $\Phi_{m,L(R)}$ are mostly localized at the left(right)
side of the trap. These modes can be built from $\Phi_{\pm}$ obtained from the 
GP equations, 
$\Phi_{m,L}={1\over \sqrt{2}}(\Phi_{m,+} + \Phi_{m,-})$, 
$\Phi_{m,R}={1\over \sqrt{2}}(\Phi_{m,+} - \Phi_{m,-})$ 
with $\Phi_{m,\pm} (\vec{r})=\pm \Phi_{m,\pm} (-\vec{r})$. 
The complex components are normalized as, 
$
\Psi_{m,L}=\sqrt{N_{m,L}(t)} e^{\imath \phi_{m,L}}$ and 
$\Psi_{m,R}=\sqrt{N_{m,R}(t)} e^{\imath \phi_{m,R}}$. Therefore, for this 
case, $z_m=  (N_{m,L}-N_{m,R})/N_m$, and
$\delta\phi_m=\phi_{m,R}-\phi_{m,L}$ $\forall m$.

As a first step we consider the so-called standard two-mode, which 
implies that all the overlapping integrals involving products of 
$L$ and $R$ modes of any two components are neglected. This approximation 
is expected to yield essentially the correct physics
as it was for the case of the scalar condensate or binary 
mixtures~\cite{ananikian,2component}.

We take the following assumptions: zero total magnetization, 
small imbalances, $z_m$, small phase differences, 
$\delta\phi_m$,  and small 
$\delta\phi_{L(R)}\equiv 2\phi_{0,L(R)}-\phi_{+1,L(R)}-\phi_{-1,L(R)}$. Also 
we take equal modes for the three states 
($\Phi_{L(R)}\equiv \Phi_{m,L(R)} \;\forall m$), which is fully justified 
at zero magnetic field (it corresponds to the single mode approximation 
at each well).

In such conditions one can prove that the total population of the 
different components, $N_m(t)$, fully decouples from the 
Josephson tunneling dynamics. The time evolution of $N_0$ is 
given by, 
\beqa
\ddot{N}_0(t)&=& 
-
4U_2^2 
N_0(t)(N-N_0(t)) (N_0(t)-N/2)
\label{eq:two0}
\eeqa
with $\hbar U_2= c_2 \int d\vec{r}
\phi_{L(R)}^4(\vec{r})$, where one can indistinctly use the left or the 
right mode. The other two follow: $N_{\pm1}(t)=(N-N_0(t))/2$. If $N_0(t)\sim
N/2$, the behavior of $N_0$ becomes sinusoidal, 
$N_0(t)=N/2 + (N_0(0)-N/2)\cos(\omega_T t)$, where we have defined 
the ``population transfer frequency'', $\omega_T=N U_2$. Eq.~(\ref{eq:two0}) 
gives an excellent agreement compared to the full GP results. With 
the considered conditions no damping is observed in the evolution of 
the populations.

The system of equations governing the dynamics of the population imbalances, 
$z_m$, and phase differences, $\delta\phi_m$, reads:
\beqa
\dot{z}_{\pm 1} &=& 
-\omega_r \delta\phi_{\pm 1} 
- (N_0 / 2)\;U_2 (\delta\phi +z_{\pm 1} \Delta\phi) \,,
 \nonumber \\
\dot{z}_{0} &=& 
-\omega_r \delta\phi_0 
+ \bar{N}\;U_2 (\delta\phi + z_0 \Delta\phi) \,, \nonumber \\
\delta\dot{\phi}_{\pm 1} &=&  
U (\bar{N} z_{\pm 1} + N_0 z_0) + U'\bar{N} z_{\mp 1} 
\nonumber \\
&&+\omega_r z_{\pm 1} +  U_2 {N_0  \over 2} (2z_0-z_{\pm 1}+z_{\mp 1}) \,,
\nonumber \\
\delta\dot{\phi}_0 &=& (U+U_2) \bar{N} (z_{-1} + z_{+1}) 
+ U_0N_0z_0 +\omega_r z_0 \nonumber \\
\Delta\dot{\phi} &=& 8 (N_0-N/2) U_2 \,,
\eeqa
where $\delta\phi=\delta\phi_L-\delta\phi_R$,
$\Delta\phi=\delta\phi_L+\delta\phi_R$, 
$\bar{N}\equiv N_{+1}=N_{-1}=(N-N_0)/2$, 
$\hbar U_0=c_0 \int d\vec{r} \phi_{L(R)}^4(\vec{r})$, 
$U=U_0+U_2$, $U'=U_0-U_2$, 
$K=-\int d\vec{r} ({\hbar^2/(2M)} \nabla\Phi_{L} \cdot \nabla\Phi_{R}
+\Phi_{L} \,V\,\Phi_{R})$, and $\omega_r=2K/\hbar$, is the Rabi frequency.

From the ground and first excited states of the system 
computed numerically, see Fig.~\ref{fig1}, we build the 
left and right modes as explained above and 
compute the microscopic parameters entering in the two-mode description. The 
resulting values are: $\omega_r=0.00386$ KHz, $NU_0=26.604$ KHz and 
$NU_2=0.12366$ KHz. This completely fixes from a microscopic level 
the parameters used in the two mode description.

\begin{table}[b]
\caption{Conditions of the different full spinor GP 
simulations,  Eqs.(~\ref{eq:gp}). $\delta\phi_m(0)=0$ in all cases. \label{tab:1}}
\begin{tabular}{l|r|r|r|r|c}
\hline
Sim  &   $N_0(0)/N$   &  $z_{-1}(0)$ &  $z_{0}(0)$ &   $z_{+1}(0)$  &
Transfer  \\
\hline
I      &     0.4        &   0.005   &  0.005      &    0.005      & YES  \\
IIa(b) &     0.6        &   0.010    &  0.000        &     0.020      &  YES(NO)\\
IIIa(b)   &      0.6        &   0.000     &  0.010       & 0.000         & YES(NO)\\
IVa(b)    &     0.6        &   0.010  &  0.000      &  $-$0.010      & YES(NO)\\
\end{tabular}
\end{table}

\begin{figure}[t]
\includegraphics[width=0.99\columnwidth, angle=0, clip=true]{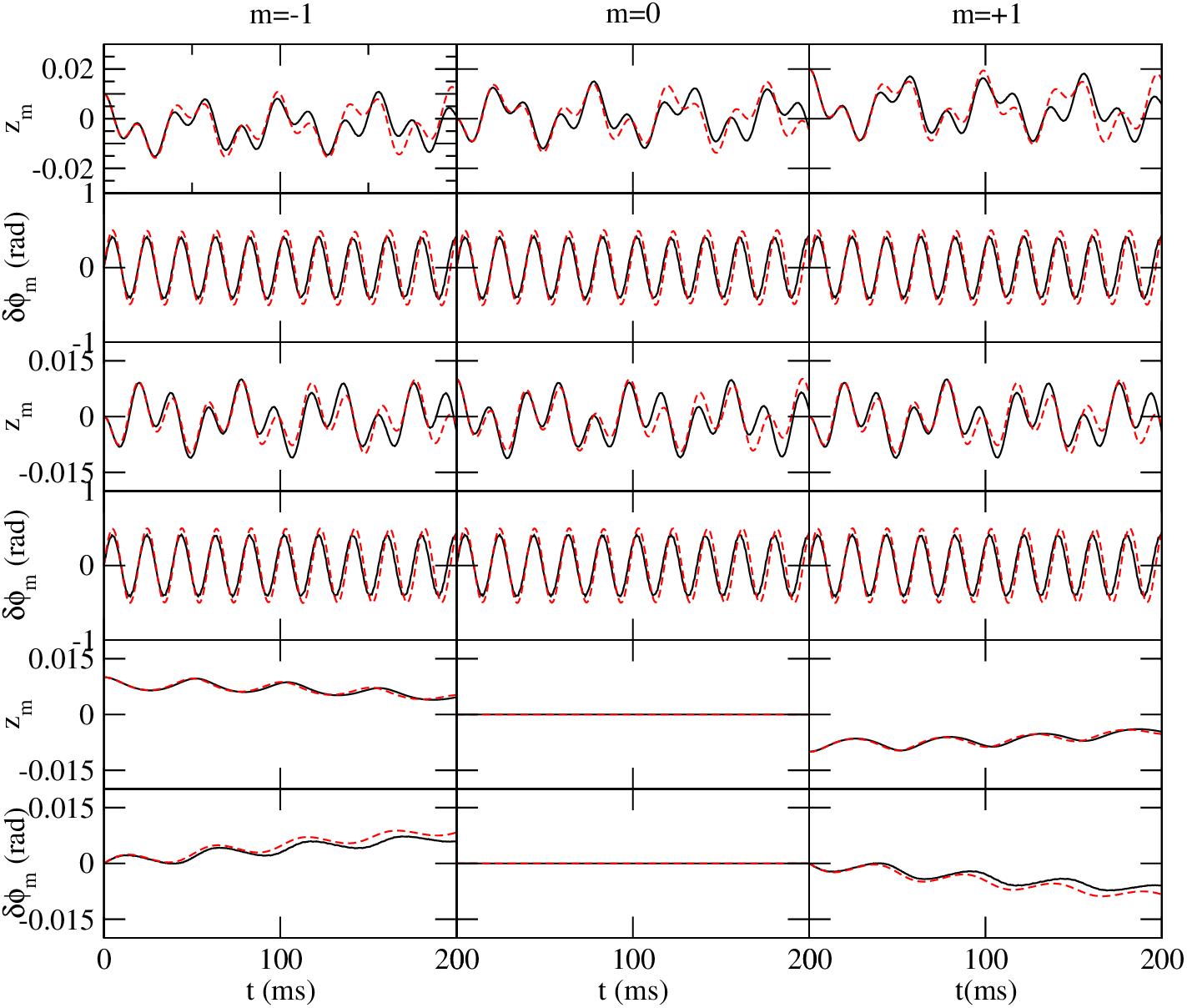}
\caption[]{Full simulation of Eqs.~(\ref{eq:gp}) and two-mode analysis 
of some cases listed in Table~\ref{tab:1}. The first/second, third/fourth 
and fifth/sixth rows correspond 
to simulations IIa, IIIa, IVa, respectively. Solid lines 
correspond to the GP simulations. Dashed lines depict two-mode results with 
the parameters computed microscopically as described in the text.
In most cases the two lines in each panel are almost indistinguishable.}
\label{fig2}
\end{figure}

First let us consider the simplest full GP simulation, listed as I in 
Table~\ref{tab:1}. This consists of the three components starting from the 
same initial population imbalances and basically gives a similar Josephson 
tunneling behavior for the three components. As can be seen in Fig.~\ref{fig1}
the Josephson regime is fully identified on the coupled behavior of $z_m$ and
$\delta\phi_m$. Together with the Josephson oscillation there is a 
transfer of population between the 
three different states, see panel (b) of Fig.~\ref{fig1}. As discussed above 
the population transfer dynamics 
decouples from the Josephson tunneling in this regime and thus allows to 
clearly identify the value of $NU_2$, which is of course directly 
linked to $c_2$. The agreement between the two-mode and the full 
GP simulation is remarkable as can be seen in Fig.~\ref{fig1}. 
Taking into account that for $^{87}$Rb $|c_2|<<c_0$ and therefore 
$U_2N<<U_0N$, it is easy to prove from the above two-mode equations 
that, for this case, the behavior of the imbalance of all the components follows: 
$\ddot{z}_m=-\omega_J^2 z_m$ with $\omega_J=\omega_r
\sqrt{1+NU_0/\omega_r}$. Which corresponds to the Josephson frequency of a scalar 
condensate completely decoupled from the population transfer~\cite{Smerzi97}. 
Therefore, the Josephson tunneling is directly related to the spin 
independent coupling, proportional to $U_0$. 

\begin{figure}[t]
\includegraphics[width=0.95\columnwidth, 
angle=0, clip=true]{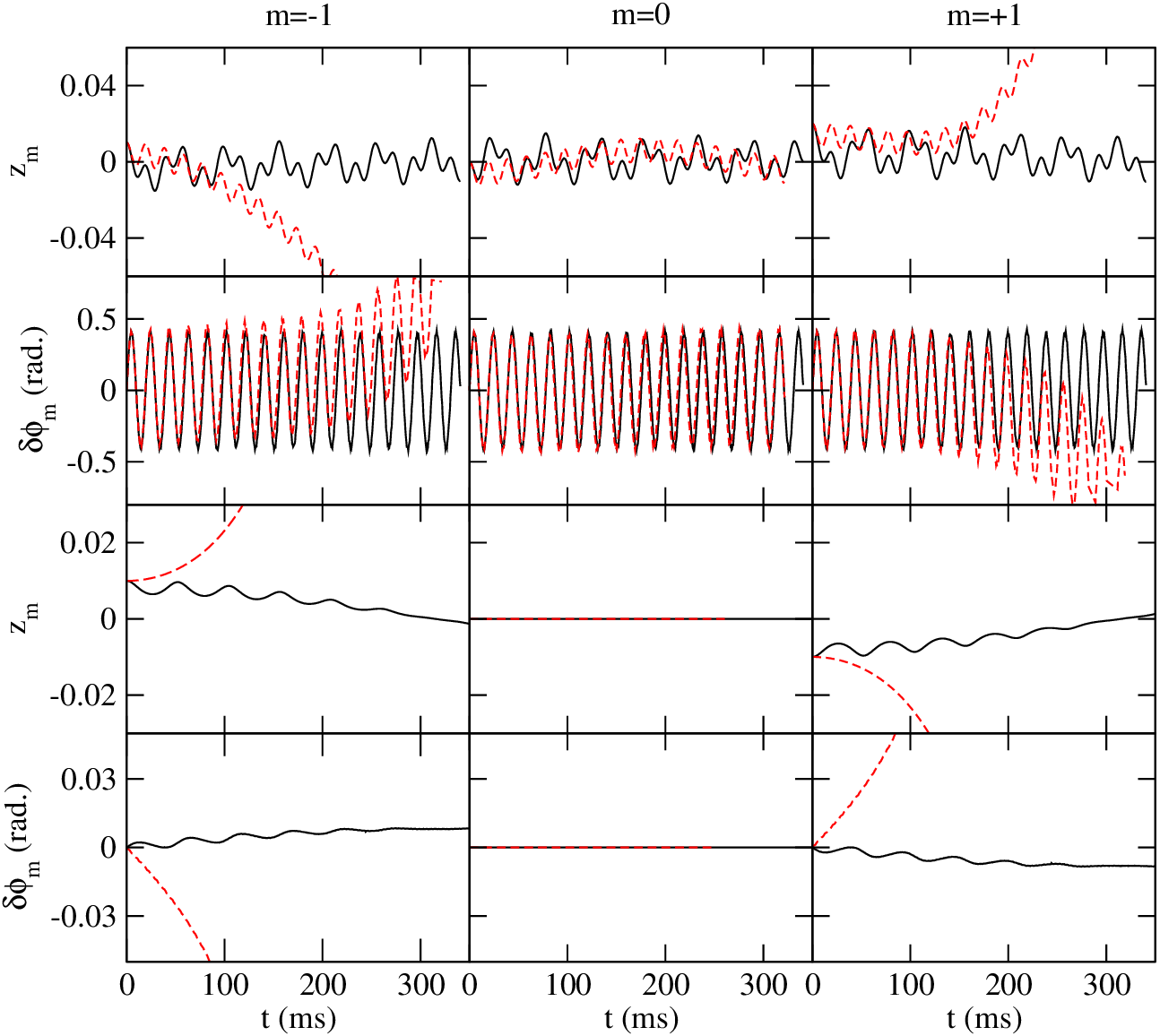}
\caption[]{The first/second, and third/fourth rows correspond 
to simulations IIa(b) and IVa(b) described in Table~\ref{tab:1}. 
Solid (black) lines correspond to IIa and 
IVa while dashed (red) lines stand for IIb and IVb, which do not 
include the population transfer terms.}
\label{fig234}
\end{figure}

Now we consider three distinct cases: IIa, IIIa, IVa. As listed in 
Table~\ref{tab:1} they correspond to different initial population 
imbalances for the three components and to a different initial number of 
atoms populating each sublevel from the one used in I. In figure~\ref{fig2} 
we show the results of the full GP simulations (solid lines). Runs IIa and IIIa produce
essentially Josephson tunneling dynamics modulated by a longer oscillation. 
Simulation IVa, produces a much longer tunneling, the 
$\pm 1$ components remain mostly on their original side of the trap while the 
$0$ one remains mostly balanced. In the first two cases the oscillations of the 
phase differences are fully characterized by $\omega_J$.
In the same figure, and almost 
indistinguishable from the full GP results, we present the predictions of 
the two-mode model. 

As mentioned above the population transfer dynamics fully decouples from the 
Josephson tunneling of the three components in the considered conditions. Its 
counterpart is however not true, the Josephson dynamics 
gets affected by the population transfer as we will discuss in the following.

\begin{figure}[t]
\includegraphics[width=0.95\columnwidth, 
angle=0, clip=true]{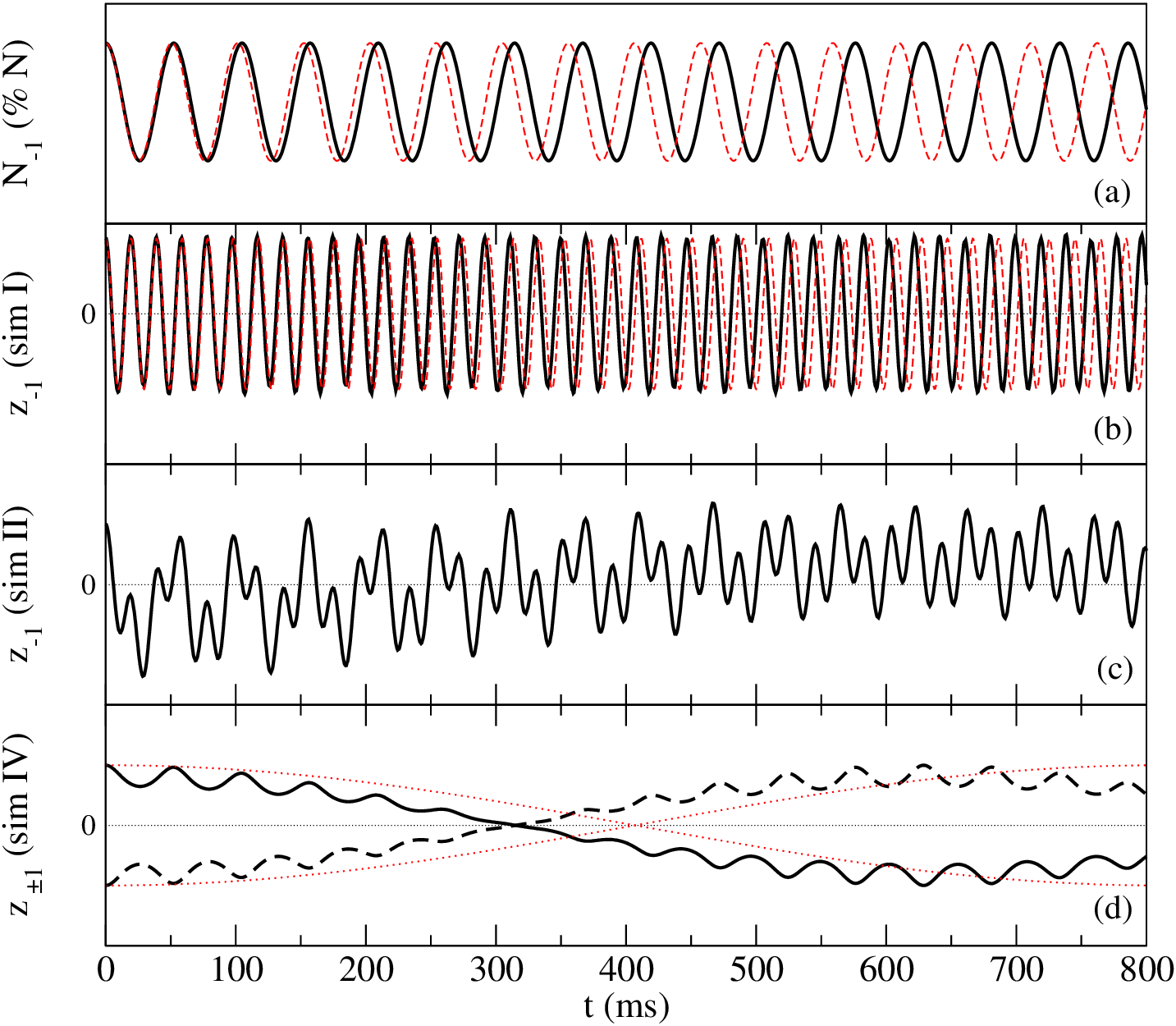}
\caption[]{Figure which shows the frequencies entering in the
  problem. (a) Time evolution of the number of atoms populating the 
$m=-1$ sublevel in simulation I, solid line. The dashed line depicts
a $\cos(\omega_T t)$ which is the two-mode prediction for $N_0\sim N/2$. 
(b) Full GP evolution for $z_{-1}$ of simulation I, solid line. The dashed line shows 
a $\cos(\omega_J t)$ behavior, clearly identifying the Josephson time scale.  
(c) Full GP evolution of $z_{-1}$ for simulation II. The dynamics is governed 
by $(\omega_T,\omega_J)$. (d) The solid black (red-dashed) line corresponds to the GP 
evolution of $z_{-1}$ ($z_{+1}$) of simulation IV. The dotted lines follow a $\cos(\omega_r t)$, 
which drives the long-time scale of the problem. The scales in the vertical 
axes are not shown for 
clarity.  
\label{fig:freq}}
\end{figure}

To clearly see the effect of the population transfer terms on top of the 
Josephson tunneling dynamics we consider the same configurations, labeled as
``a'', but without the population transfer terms, ``b''. The two-mode model, 
without the corresponding transfer terms, also reproduces the dynamics of 
the ``b'' runs. In Fig.~\ref{fig234} we depict
in all cases a comparison between the full GP solution 
and the same case but neglecting the population transfer term.

The effects of population transfer are clearly seen on the 
evolution of $z_m$. In simulation II, which 
has $z_0(0)=0$ it is observed that the long oscillation which modulates 
the full runs, $\omega_T$, is not present when we switch off the transfer 
term. Instead the population imbalance shows a Josephson-like tunneling 
oscillation which for $t \sim 100$ ms looses the small $z_m$ regime.  
Therefore, the transfer term tends to stabilize the Josephson-like 
behavior over longer periods of time. The absence of transfer
of populations does not show up on the behavior of the phase 
difference, as can be seen in Fig.~\ref{fig234}, which mostly follows 
the same evolution as for the GP equations with the transfer term.

As in the case of binary mixtures~\cite{bjd09}, taking opposite 
initial imbalances for the $m=\pm 1$ components enhances the Rabi 
like oscillation and cancels the Josephson one. Simulation IV 
corresponds to such a case, with $z_{-1}(0)=-z_{+1}(0)$ and 
$z_0(0)=0$. The Rabi oscillation gives rise to a long tunneling 
behavior but in this case modulated by the $\omega_T$ oscillation, 
as can be seen in Fig.~\ref{fig234} and in the lowest panel of Fig.~\ref{fig:freq}. If we 
switch off the transfer term the $\omega_T$ oscillation 
disappears and the limit of small $z$ and $\delta\phi$ becomes unstable.

Finally, Fig.~\ref{fig:freq} summarizes the relevant frequencies which 
enter in the interplay between Josephson tunneling and population transfer dynamics 
in the considered regime. The first panel isolates $\omega_T=NU_2$, governing 
the transfer of populations, whereas the second 
one shows $\omega_J$, which sets the fast behavior of the imbalances. 
The third panel shows $z_{-1}$ from simulation II,
which is dominated by ($\omega_T$,$\omega_J$) and the fourth one shows
both $z_{\pm 1}$ from simulation IV, that are dominated by two 
frequencies ($\omega_T$, $\omega_r$).

The ability to perform an experiment with spinor $F=1$ BEC in the 
conditions considered in this work would present for the first time the 
combined effects of Josephson tunneling phenomena 
and the transfer of population between different Zeeman components of a 
spinor condensate: the decoupling of the exchange of populations from 
the Josephson dynamics, the identification of the different time scales and 
the role of the population transfer in the stability of the Josephson 
oscillations. In addition, a precise measurement of the population imbalances and 
global populations of the three species would provide an alternative access to
the microscopic properties of the atom-atom interactions.

We thank Joan Martorell for a careful reading of the manuscript 
and motivating discussions. B.J-D. is 
supported by a CPAN CSD 2007-0042 contract, Consolider 
Ingenio 2010. This work is also supported by the Grants 
No. FIS2008-01661, FIS2008-00421 and 2009SGR-1289 from 
Generalitat de Catalunya.

\end{document}